\documentstyle[aps,psfig]{revtex}
\begin{document}
\baselineskip=24pt
\begin{center}
\bf{FINITE TEMPERATURE PROPERTIES OF SO(3) LATTICE GAUGE } \\
\bf{THEORY AND THEIR IMPLICATIONS FOR THE CONTINUUM THEORY}

\vspace{1cm}

\rm{SRINATH CHELUVARAJA}\footnote[1]{e-mail:srinath@imsc.ernet.in} \\
 and \\
\rm{H.S. SHARATHCHANDRA}\footnote[2]{e-mail:sharat@imsc.ernet.in} \\
The Institute of Mathematical Sciences \\
Madras - 600 113, INDIA\\
\end{center}
\vspace{1cm}

\noindent{\bf{ABSTRACT}}\\
It is shown that $SO(3)$ lattice gauge theory on finite size lattices has
metastable states related to the ground states of both the bulk transition
and the finite temperature transition. The Polyakov line variable in the
adjoint representation of $SU(2)$ is used to trace the origin of these
metastable states. It is also argued that a second order finite temperature
transition in the continuum theory is not inconsistent with the first
order transition in $SO(3)$ lattice gauge theory and the absence
of a $Z(2)$ global symmetry.

\vspace{0.5cm}
\begin{flushleft}
\end{flushleft}

\newpage

There are many arguments that the Yang Mills (Y-M) theory undergoes a second
order phase transition into a deconfined phase at finite temperature. The
strong coupling limit of $SU(2)$ lattice gauge theory at $T\ne 0$ can be
rewritten as a spin model with a global $Z(2)$ invariance \cite{sus}. At high
temperatures, this symmetry is spontaneously broken. Monte Carlo
simulations of $SU(2)$ lattice gauge theory (LGT) have confirmed that
this result is not an artifact of the strong coupling limit and further
indicate that the transition is of second order \cite{mac}. The order parameter
for the transition is the Wilson-Polyakov line  \\
\begin{eqnarray*}
L_{f}=Tr_{f}\ P\ exp\ i\int_{0}^{\beta}dx_{4}A_{0}(\vec x,x_{4})
\end{eqnarray*}
in the language of the continuum theory. Here the subscript $f$ indicates
that the trace is taken in the fundamental representation.
The $Z(2)$ group is related to the centre of the gauge group $SU(2)$
and its non trivial element has the action \\
\begin{eqnarray*}
Z_{2}:L_{f}\rightarrow -L_{f}
\end{eqnarray*}
It has been argued that this transition should be in the universality
class of the 3-d Ising model \cite{yaffe}. This is also supported by simulations.

Simulations of LGT's with mixed actions have cast doubt on the above
picture \cite{gav}. The Bhanot-Creutz \cite{creu} model has the action \\
\begin{eqnarray*}
 \sum_{p}\left ( (\beta /2) Tr_{f}U(p) + (\beta_{a}/3) Tr_{a}U(p)\right )
\end{eqnarray*}
which is a sum of the actions in the fundamental and the adjoint
representations. This LGT exhibits distinct phases even at $T=0$ (Fig.1).
This bulk transition has been traced to a condensation of $Z_{2}$
monopoles in the small $\beta_{a}$ region \cite{hall2}. There is a line
of first 
order transitions with an endpoint of second order transition which
stops short of intersecting the $\beta_{a}=0$ line.
Study of this model \cite{gav} at $T\ne 0$ seems to suggest that the line of
finite temperature transitions is continuously connected to the line of
bulk first order transitions. This has led to a variety of contradictory
interpretations in the literature. (i) It has been proposed that
the bulk transition may not be really present and the effect is simply that of
the finite temperature transition in simulations with finite lattices.
(ii) The opposite scenario that the Y-M theory may not have a finite 
temperature transition at all has also been considered.(iii) The possibility
that the Y-M theory has a first rather than a second order transition is also
supposed.(iv) On the other hand, some studies, especially of related
models with $SU(3)$ and $SU(4)$ groups \cite{mix} suggest that the transitions 
are distinct and their separation can be noticed with larger lattices.\\

In this paper we make a careful study of $SO(3)$ LGT (i.e. the $\beta =0$
line of the Creutz-Bhanot action) at $T\ne 0$, in order to arrive at a
clear picture of the situation.The aim of our study is to get a clue
to the nature of the high temperature phase. It is to be noted that the
$SO(3)$ LGT does not have a global $Z_{2}$ symmetry of the corresponding
$SU(2)$ LGT, which plays a crucial role \cite{sus,mac,yaffe}
 in the finite temperature
transition. Also the finite temperature transition is of first order,if
it is there at all. Nevertheles, we expect the two theories to  have the
same continuum limit. (We comment more on this later). It is therefore to
be wondered whether the global $Z_2$ symmetry and universality
arguments are relevant for  the continuum theory. There are also other
ways in which a study of $SO(3)$ LGT can yield us a clue to the nature
of the high temperature phase. For instance, if the transition in $SO(3)$
LGT on symmetric lattices is really related to the finite temperature
transition, we could conclude that the $Z_2$ monopoles are responsible
for this effect of heating the system. 

Most of the previous investigations have generally 
measured $\langle |L_{f}| \rangle$
instead of $\langle L_{f} \rangle$
to distinguish the high temperature phase.We have realized that it is
far more instructive and not more difficult to measure $\langle L
\rangle $
itself in
order to understand the nature of the phases. It is also
very useful to measure this order parameter in each Monte-Carlo sweep
and follow its evolution. For the $SU(2)$ LGT
in the high temperature phase $\langle L_f \rangle$
preferably settles to a positive value with a cold start. With a hot start
it is equally likely to go to this state or the one with opposite sign
of $\langle L_f \rangle$ \cite{mac}.
Thus two distinct ground states 
related by the $Z_2$ transformation are explicitly seen. In contrast, in the
low temperature phase, $\langle L_f \rangle$ is consistent with zero.

We have repeated a similar measurement for the adjoint Polyakov line
$\langle L_a \rangle$ in $SO(3)$ LGT. There is no reason to regard this
as an order parameter for the $SO(3)$ theory. $\langle L_a \rangle$
measures $\exp (-\beta F_{a})$ where $F_a$ is the free energy of a
static quark in the adjoint representation of $SU(2)$. Even in the
presumed confinement phase, such a quark can form a colour singlet bound
state with the gluon, also in the adjoint representation, and thus give
$F_{a}< \infty$. Nevertheless, we have found that $\langle L_{a} \rangle$
serves as a good observable to distinguish various phases.

We find that $\langle L_{a} \rangle$ is very small in the low 
temperature phase (Fig.2). It is, in fact, consistent with zero within errors.
However there is no reason to expect it to be exactly zero, in contrast
to $\langle L_{f} \rangle$ in the $SU(2)$ theory. In the strong coupling
expansion, we get $\langle L_{a} \rangle = O(\beta_{a}^{4N_{\tau}})$  which
is small for the couplings and lattice sizes we are using. This is
presumably the reason for the smallness of $\langle L_{a} \rangle $.
$\langle L_{a} \rangle $ has a far more interesting behaviour in the
high temperature phase (Fig.2). With a cold start it settles to a positive value
whereas with a hot start it settles to a negative value, distinct in
magnitude (Fig.3). Thus it appears that there are two ground states for $SO(3)$
LGT at high temperatures, just as for the $SU(2)$ case. But, there is no
indication of a $Z_2$ symmetry connecting the two ground states. We have
measured the value of the free energy in the two ground states and find that
they are equal within errors. Therefore it appears that there are two 
distinct
and degenerate ground states.

We need to find the raison d' etre of the two ground states. Note that \\
\begin{eqnarray*}
\langle L_{f} \rangle = 2\cos (\theta /2) \\
\langle L_{a} \rangle = 1 + 2\cos (\theta)
\end{eqnarray*}
and therefore have ranges $L_{f}=[-2,2]$ and $L_{a}=[-1,3]$, where
$\exp \pm i(\theta /2)$ are the two eigenvalues of the $SU(2)$ matrix \\
\begin{eqnarray*}
U=Tr_{f}\ P\ exp\ i\int_{0}^{\beta}dx_{4}A_{0}(\vec x,x_{4})
\end{eqnarray*}
and $\theta$ has the range $[0,4\pi]$. It is to be noted that since
U transforms under local gauge transformations as \\
\begin{eqnarray*}
  U\rightarrow V(\vec x,0) U V^{\dagger}(\vec x,0)
\end{eqnarray*}
these eigenvalues are gauge invariant observables.

The $Z_2$ global symmetry group of the $SU(2)$ LGT corresponds to 
$\theta \rightarrow \theta + 2\pi$. The fact that $\langle L_{f} \rangle$
is non zero and has either of two equal and opposite values can be
interpreted as follows. The high temperature phase of $SU(2)$ LGT has
configurations $\theta (mod 4\pi)$ which are dominantly in one or the
other sector, $[0,2\pi]$ or $[2\pi,4\pi]$. This bifurcation of the
configurations is explicitly seen \cite{mac,nex}. In contrast,
in the low temperature phase, the configurations are symmetrical \cite{nex}
about 
$\langle \cos \theta/2 \rangle = 0$.

The variable $\langle L_{a} \rangle$ is insensitive to the $Z_2$
transformation. Therefore, if we were to probe the $SU(2)$ LGT using
it,  we would see just one ground state in the high temperature phase. 
Moreover
this ground state would correspond to values in the region $[0,3]$ since
$\langle \cos \theta \rangle$ is dominantly positive for such
configurations.

This now provides an explanation for the ground state   in the domain
$[0,3]$ of the $SO(3)$ LGT at high temperatures, one that is preferred in
the cold start. It is just the high temperature phase of the corresponding
$SU(2)$ LGT. Histograms show (Fig.4a) that the configurations are peaked
around positive values of $\cos \theta$ for this ground state.

We have to
now account for the other ground state for which $\langle L_a \rangle$
is negative (Fig.4b).
We now argue that this ground state is really the vaccuum of the large
$\beta_{a}$ phase of the bulk transition in $SO(3)$ LGT. Strictly
speaking the order parameters 
$ \langle L_f \rangle $ and $\langle L_a \rangle $ are zero at $T=0$ because
then the lattice extends indefinitely in the $x_4$ direction. However,
we are working with lattices of finite extent in both the $x_4$ and spatial
directions. Therefore $\langle L_f \rangle $ 
or $\langle L_a \rangle $ need not be strictly vanishing. In
a finite lattice, phase transitions are impossible, in principle. At the
most we can see metastable states and tunnelling between them. As we
increase the lattice size, the system prefers to be in one of these
metastable states longer and longer. In the thermodynamic limit, the
system spends all the time in (one of equivalent) ground state(s).

The lattice of unequal extent in the $x_4$ and spatial directions can
be interpreted equally as a finite volume version of a LGT with
anisotropic couplings at $T=0$ or as a system at a non zero temperature.
Therefore we would expect to find the metastable states corresponding to
the large $\beta_a$ phase of the bulk transition in $SO(3)$ LGT
and also the high temperature phase. Even though these states are not
degenerate, with a skewed initial configuration the system may spend a long
time in the state with a higher free energy. Indeed, the way to pin down
the true ground state would be by using a hot start. Thus we would expect
that on symmetric lattices, with a hot start the system settles to the vaccuum 
of the bulk transition. On the other hand, on an asymmetric lattice the hot start would choose the finite temperature ground state.

These expectations are borne out in our simulations and support the
idea that the ground state with $\langle L_{a} \rangle $ in the interval 
$[0,3]$ corresponds
to the high temperature phase and the one with $\langle L_{a} \rangle$   
in the interval $[-1,0]$
corresponds to the bulk phase of $SO(3)$. In Fig.5 we notice that
with a hot start, on the asymmetric lattice,the system
settles into the phase with $\langle L_{a} \rangle$ positive.
On the other hand, in case of a
symmetric lattice the system prefers to be in the phase with
$\langle L_{a} \rangle$  negative and small. On the large $8^{4}$ lattice this
preference
is even more marked \cite{nex}. We have found further 
evidence \cite{nex} to support this picture.

  Our investigation confirms that the finite temperature transition is
neither a lattice artifact nor an artifact of the action being 
used. $SO(3)$ LGT has
a finite temperature transition analogous to that in the $SU(2)$ LGT, even
though there is no global $Z_2$ symmetry to provide an order parameter in the
strict sense. Instead, it is the bunching of configurations around a preferred
domain of $\theta$ (the phase of the eigenvalues of the unitary matrix)
which characterizes the high temperature phase. This is an important change
in the
characterization of this phase because it can be used in the presence of
matter fields also. We use it elsewhere \cite{ani} to characterize the nature
of the high
temperature phase of QCD and to obtain a phenomenological model for it.

We feel the need to address two more questions to further justify the above
conclusions. In the Bhanot-Creutz phase diagram (Fig.1) the large $\beta$
 ($\beta_a$) phases of $SU(2)$ ($SO(3)$) theories are separated by a phase 
  boundary.
Therefore the continuum limits are taken in distinct phases. However there is
another version of the mixed action by Halliday and Schwimmer \cite{hall2} 
in which the
large $\beta$ phases of $SU(2)$ and $SO(3)$ theories are continuously
connected. Therefore the separation by a phase boundary in the Bhanot-Creutz
model must be an artifact of the action chosen.

Even if we accept that the phase with $\langle L_{a} \rangle$ in the domain
$[0,3]$ in the $SO(3)$ LGT is the high temperature phase, we have to tackle
the problem that the transition is now of first order, in contrast to the
second order transition in $SU(2)$ LGT. We need to reconcile this with its
relevance to the continuum theory. We believe that the latent heat $\Delta L$
of this first order transition scales to zero in the continuum limit being
considered. In other words, as we decrease the lattice spacing keeping the
physical $\Lambda$ parameter fixed, $\Delta L/\Lambda$ scales to zero. We
hope to check this in future simulations.

  In this paper we have shown that simulations with finite size lattices
may show metastable states related to the ground states of both bulk and
finite temperature transitions and
argued that the finite temperature transition persists independently of the
specific action used even if the global symmetry is absent. We have suggested
that a second order finite temperature transition can persist in the
continuum theory even though it is of first order in $SO(3)$ LGT.

{}
\newpage
\begin{figure}[ht]
\caption{Phase diagram for the Bhanot-Creutz action.The dashed
line indicates the finite temperature transition studied by
Gavai et.al.}
\end{figure}

\begin{figure}[ht]
\caption{$\langle L_{a} \rangle$ plotted as a function of 
$\beta_a$  on a $7^{3}\  3$ lattice.}
\end{figure}

\begin{figure}[ht]
\caption{$L_{a}$ values for hot start (unbroken lines) and 
cold start (broken lines) as a function of Monte Carlo sweeps
on a $7^3\ 3$ lattice for $\beta_a=3.5$.}
\end{figure}

\begin{figure}[ht]
\caption{$L_{a}$\ $(1+2\cos \theta )$ 
histograms for the two metastable states on a $7^3\ 3$ lattice.(a) For the
one arising from a cold start and (b) for the one arising from a hot start.
The value of the $\beta_a$ was 3.5.}
\end{figure}

\begin{figure}[ht]
\caption{$L_{a}$ plotted vs number of sweeps on a $7^4$ symmetric lattice
(solid lines) and on a $7^3\ 3$ lattice (broken lines) with a hot start
at $\beta_a = 3.5$.} 
\end{figure}
\newpage

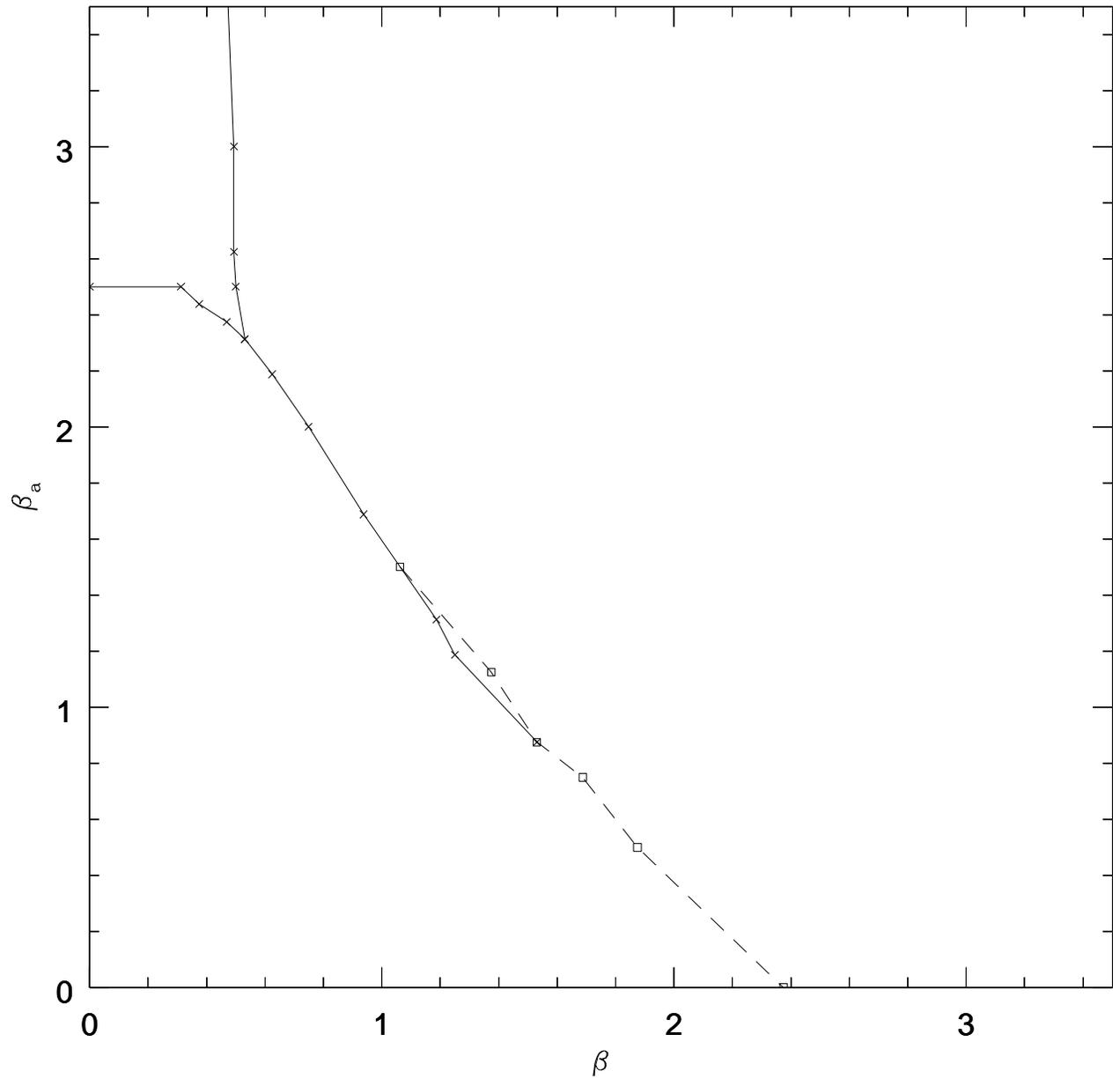
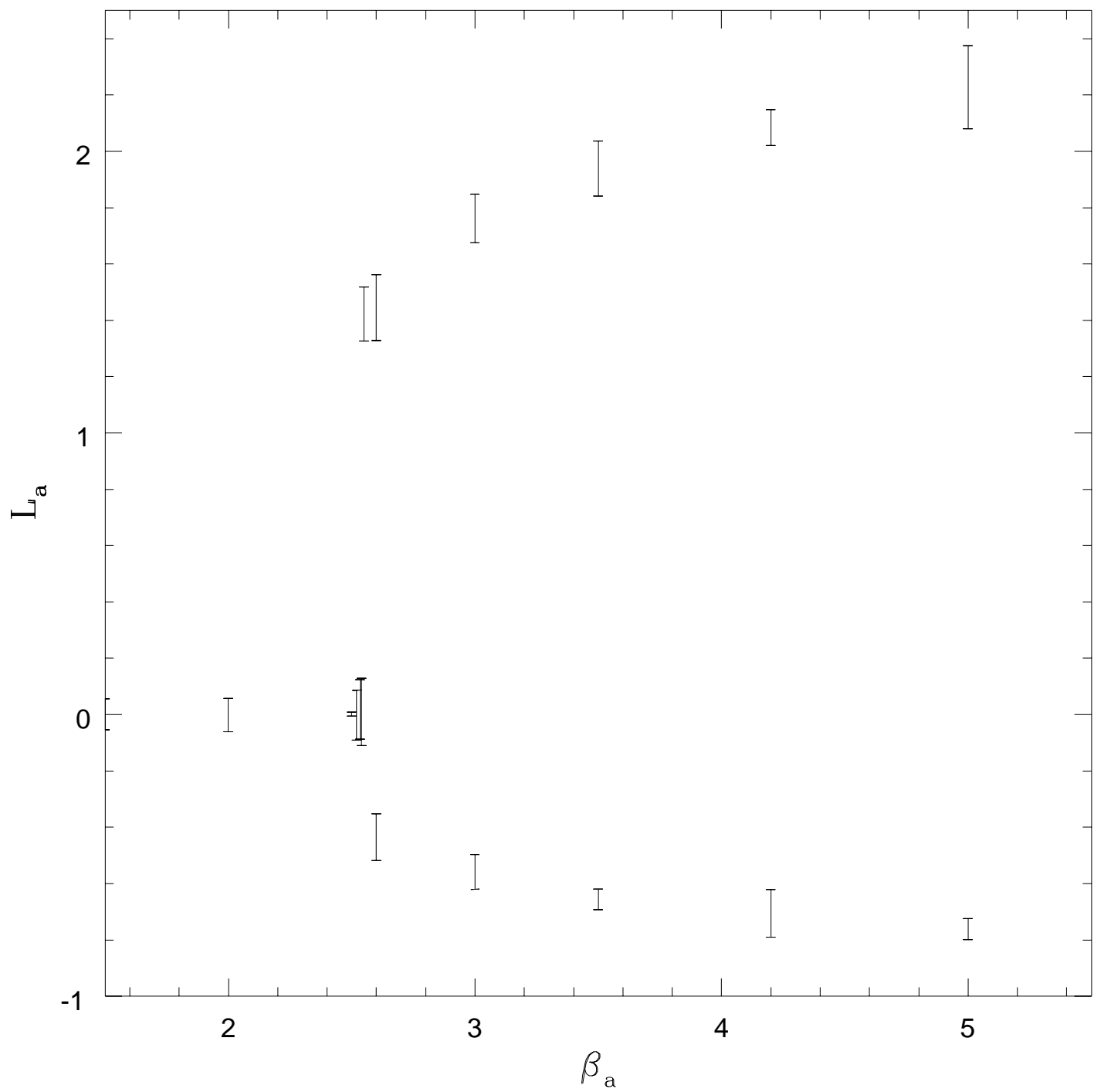
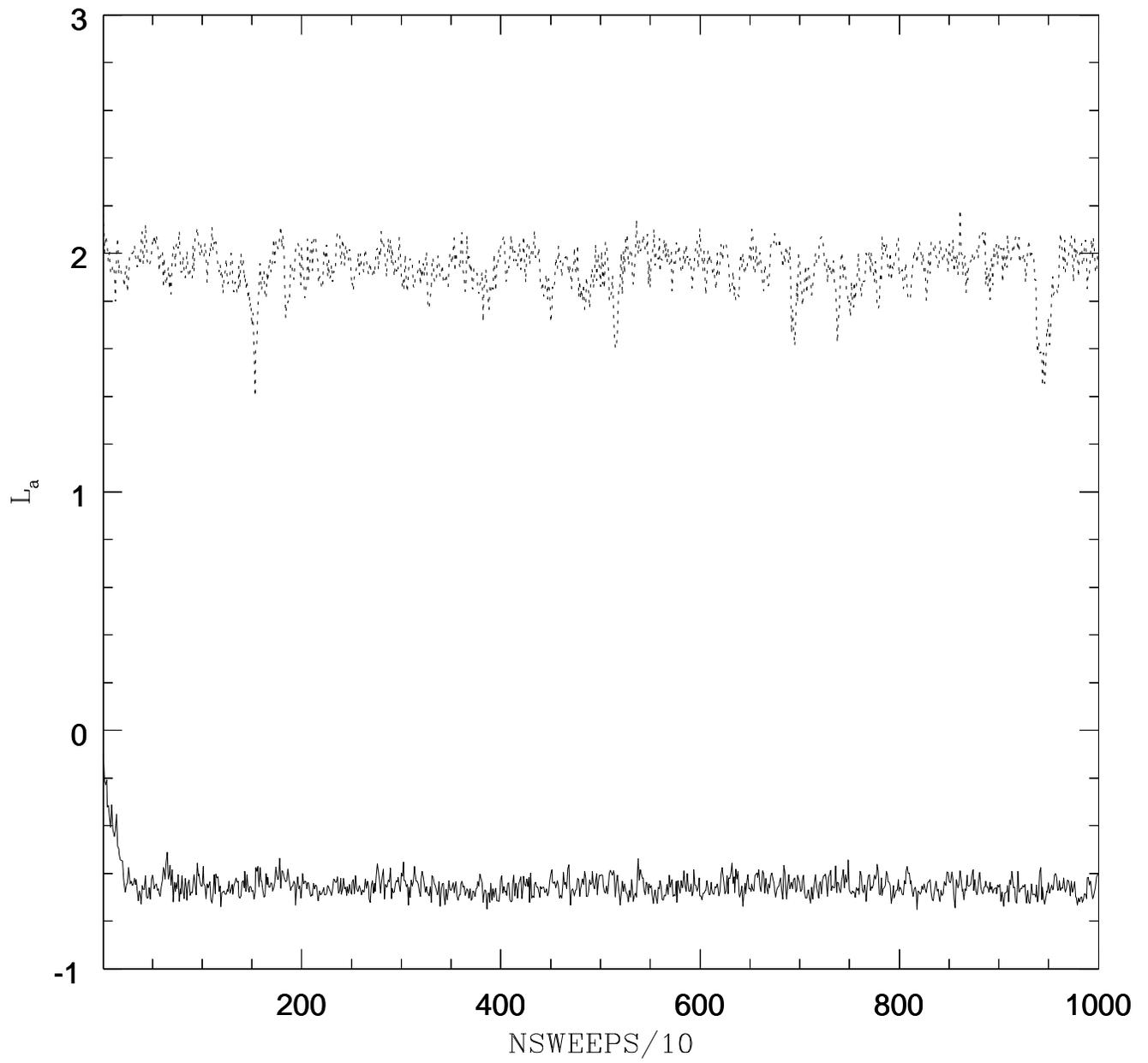
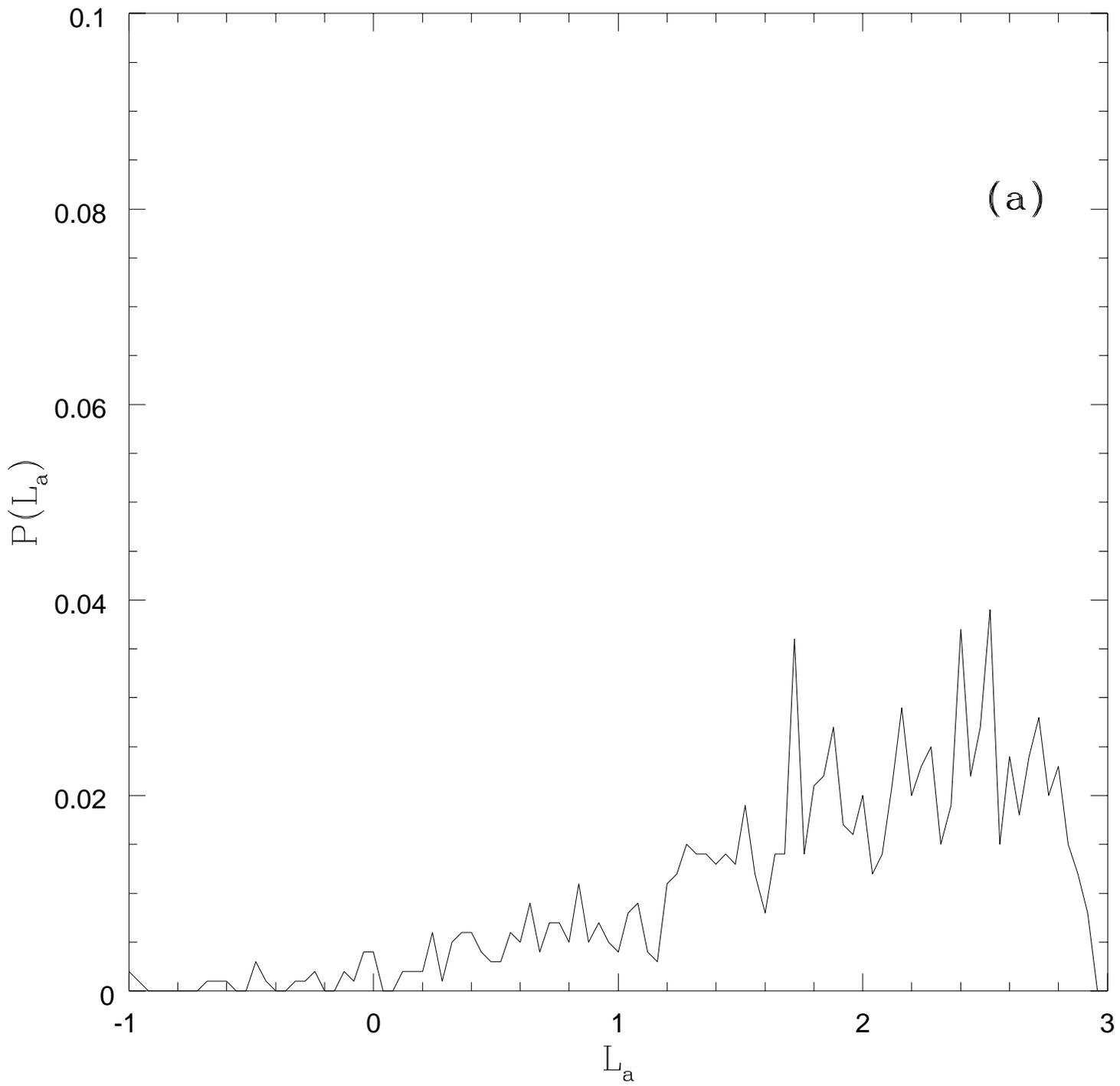
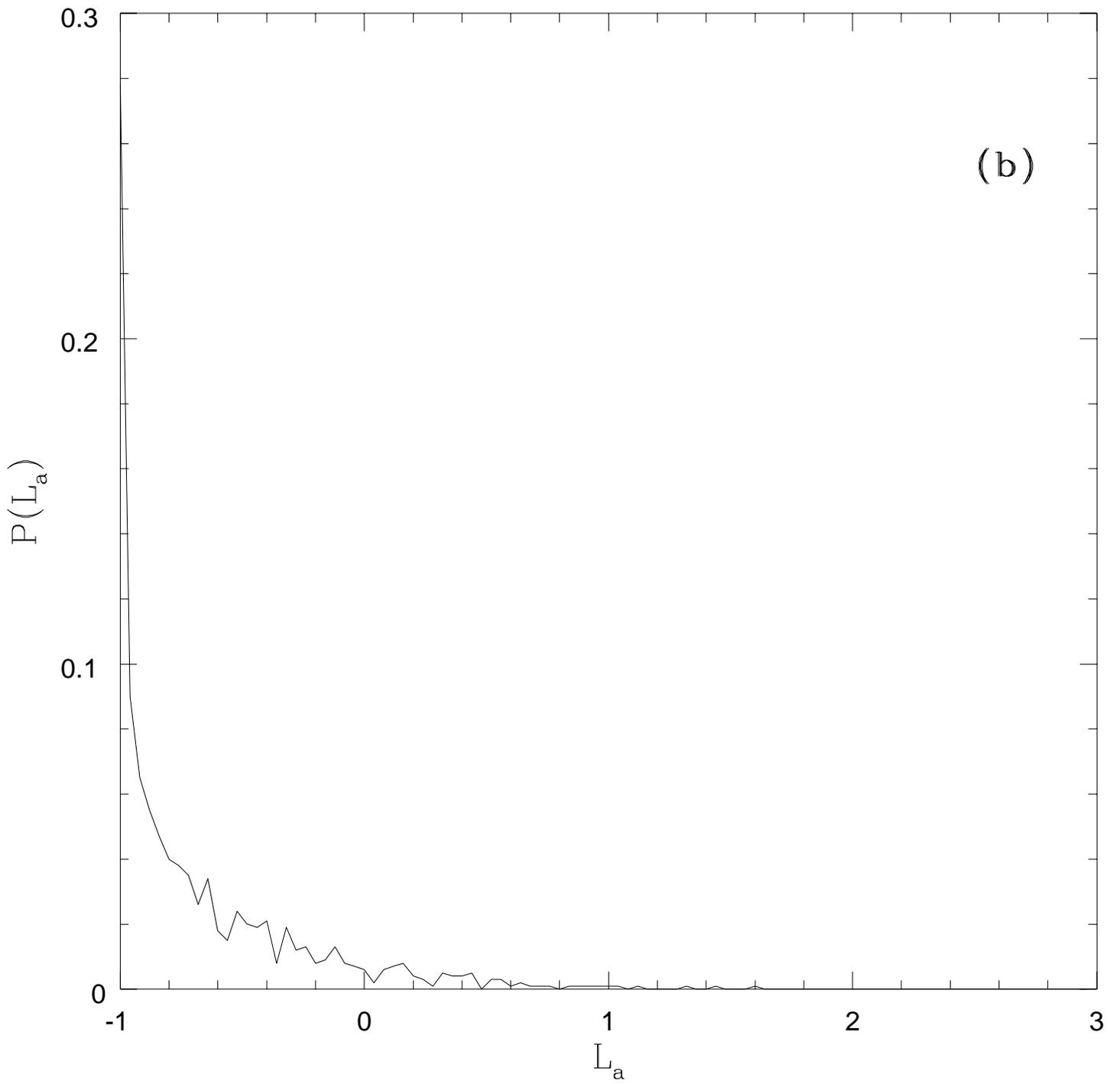
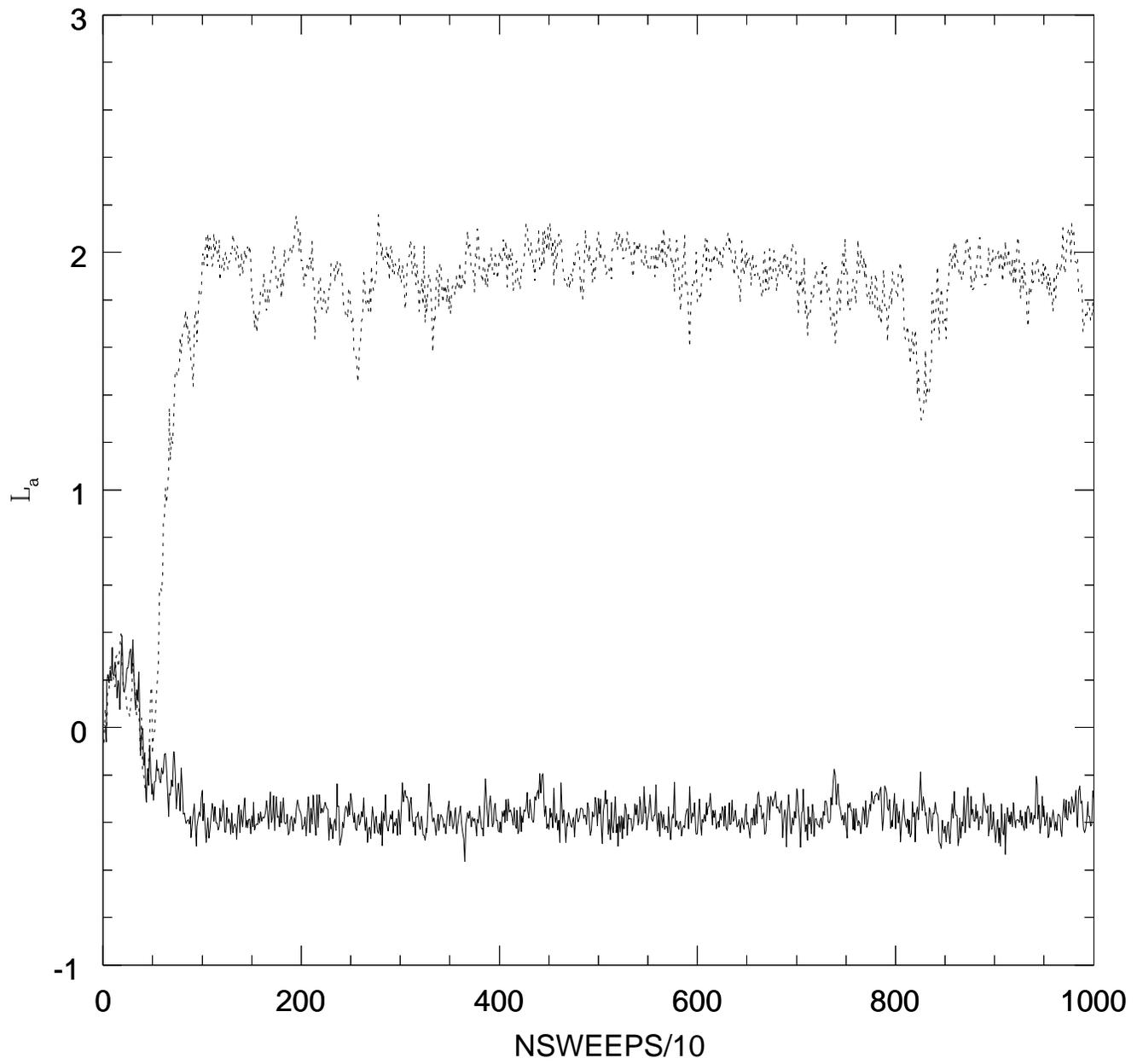
\end{document}